\documentclass[aapm,graphicx]{revtex4-1}

\draft % marks overfull lines with a black rule on the right

\usepackage[mathlines]{lineno}% Enable numbering of text and display math
\usepackage{amssymb}
\usepackage[pdftex]{graphicx}
\usepackage{multirow}

\begin{document}

%\preprint{}

\title{Optimization of a multipoint plastic scintillator dosimeter for high dose rate brachytherapy} %Title of paper



\author{Haydee M. Linares Rosales$^{1,2}$}
\author{Patricia Duguay-Drouin$^{1,2}$}
\author{Louis Archambault$^{1,2}$}
\author{Sam Beddar$^{3}$}
\author{Luc Beaulieu$^{1,2}$}

\affiliation{$^{1}$D\'epartement de physique, de g\'enie physique et d'optique et Centre de recherche sur le cancer, Universit\'e Laval, Qu\'ebec, Canada.}
\affiliation{$^{2}$D\'epartement de radio-oncologie et Axe Oncologie du CRCHU de Qu\'ebec, CHU de Qu\'ebec - Universit\'e Laval, QC, Canada.}
\affiliation{$^{3}$Department of Radiation Physics, The University of Texas MD Anderson Cancer Center, Houston, TX 77003, United States.}

\email[Corresponding author: Haydee M. Linares Rosales, ]{haydee8906@gmail.com}
\date{\today}

\begin{abstract}
\scriptsize{\textbf{Purpose:} 
This study is devoted to optimizing and characterizing the response of a multipoint plastic scintillator detector (mPSD) for application to in vivo dosimetry in HDR brachytherapy.\\

\textbf{Methods:}
An exhaustive analysis was carried out in order to obtain an optimized mPSD design that maximizes the scintillation light collection produced by the interaction of ionizing photons. More than 20 prototypes of mPSD were built and tested in order to determine the appropriate order of scintillators relative to the photodetector (distal, center or proximal), as well as their length as a function of the scintillation light emitted. The available detecting elements are the BCF-60, BCF-12 and BCF-10 scintillators (Saint Gobain Crystals, Hiram, OH, USA), separated from each other by segments of Eska GH-4001 clear optical fibers (Mitsubishi Rayon Co., Ltd., Tokyo, Japan). The contribution of each scintillator to the total spectrum was determined by irradiations in the low energy range ($<$ 120 keV). For the best mPSD design, a numerical optimization was done in order to select the optical components (dichroic mirrors, filters and photomultipliers tubes (PMTs)) that best match the light emission profile. Calculations were performed taking into account the measured scintillation spectrum and light yield, the manufacturer-reported transmission and attenuation of the optical components, and the experimentally characterized PMT noise. The optimized dosimetric system was used for HDR brachytherapy measurements. The system was independently controlled from the $^{192}$Ir source via LabVIEW and read simultaneously using an NI-DAQ board. Dose measurements as a function of distance from the source were carried out according to TG-43U1 recommendations. The system performance was quantified in terms of signal to noise ratio (SNR) and signal to background ratio (SBR). \\

\textbf{Results:} 
For best overall light-yield emission, it was determined that BCF-60 should be placed at the distal position, BCF-12 in the center and BCF-10 at the proximal position with respect to the photodetector. This configuration allowed for optimized light transmission through the collecting fiber and avoided inter-scintillator excitation and self-absorption effects. The optimal scintillator length found was of 3 mm, 6 mm, and 7 mm for BCF-10, BCF- 12 and BCF-60, respectively. The optimized luminescence system allowed for signal deconvolution using a multispectral approach, extracting the dose to each element while taking into account the Cerenkov stem effect.  Differences between the mPSD measurements and TG-43U1 remain below 5\% in the range of 0.5 cm to 6.5 cm from the source. The dosimetric system can properly differentiate the scintillation signal from the background for a wide range of dose rate conditions: the SNR was found to be above 5 for dose rates above 22 mGy/s  while the minimum SBR measured was 1.8 at 6 mGy/s.\\

\textbf{Conclusion:} 
Based on the spectral response at different conditions, an mPSD was constructed and optimized for HDR brachytherapy dosimetry. It is sensitive enough to allow multiple simultaneous measurements over a clinically useful distance range, up to 6.5 cm from the source. This study constitutes a baseline for future applications enabling real time dose measurements and source position reporting over a wide range of dose rate conditions.\\

\textbf{Keywords:}
{in vivo dosimetry, plastic scintillator, multi points plastic scintillation detector, HDR brachytherapy}}

\end{abstract}

\pacs{}% insert suggested PACS numbers in braces on next line

\maketitle %\maketitle must follow title, authors, abstract and \pacs


%\section{TEST}
%%\label{}
%\subsection{BB}
%\subsubsection{JJ}

\section{Introduction}
In brachytherapy, radioactive sources are placed at a short distance from the target. The high dose gradients near brachytherapy sources (10\% or more per millimeter for the first few centimeters from the source) provide a level of protection to healthy tissues surrounding the target. Despite the short distances involved in this modality, brachytherapy is not free from errors, which can be caused by humans (e.g., incorrect medical indication, source strength, patient identification, catheter, or applicator) or by failures in the treatment system (e.g., mechanical events) \cite{Venselaar-Compr-Brachy}. Even small errors in the source positioning can result in harmful consequences for patients. Systematic implementation of precise quality control and quality assurance protocols help to improve treatment quality, and routine use of real-time verification systems and in vivo dosimetry are even more helpful in determining whether there are deviations from the prescribed dose during treatment delivery. Performing these tasks requires a precise and accurate detector whose presence does not perturb the particle fluence and the physics interactions. Tanderup et al.\cite{Tanderup-invivo-Brachy-2013} reviewed different detectors for potential use for in vivo brachytherapy dosimetry. The selection of the appropriate dosimetric system is a compromise between different requirements and constraints of the detector as well as the application sought.

Plastic scintillator detectors (PSDs) show promise for obtaining accurate real-time radiotherapy dose measurements. Previous studies have demonstrated that PSDs can accurately measure dose in external beam radiotherapy and that they have high spatial resolution, linearity with dose, energy independence in the megavolt energy range, and water equivalence \cite{Beaulieu-Scint-Status-2013, Therriault-Temp-method-2015, Boivin-2016, Lambert-Cerenkov-2008, Beddar-Cerenkov-1992, Archambault-2006, Wootton-Temperature-2013, Guillot-toward-2010, Beddar-water-equivalent-1992-1, Beddar-water-equivalent-1992-2}. In addition, some authors have found PSDs to be feasible for use in brachytherapy applications \cite{Lambert-2006, Therriault-2011, Therriault-Phantom-2011, Liu-real-time-2012, Therriault-Validation-2017}. Despite the aforementioned advantages, PSD response is affected by the stem effect and temperature dependence. Temperature dependence was long considered to be negligible, but recent studies showed that, depending on the type of scintillator, changes on the order of 0.6\% per degree Celsius should be expected \cite{Wootton-Temperature-2013, Beddar-temp-2012}. Moreover, a non-negligible fraction of the light collected by PSDs consists of the stem effect which can be caused by two phenomena. The first phenomenon is direct excitation of the polymer chain, or fluorescence, from the plastic optical fiber guides, and the second is Cerenkov light production. Therriault-Proulx et al. \cite{Therriault-Nature-2013} found that fluorescence yield is order of magnitude less intense than Cerenkov light. The intensity of Cerenkov light emitted in the optical fiber guide is up to 2 orders of magnitude lower per millimeter than the intensity of the scintillation light produced by the scintillator. However, the optical fiber guides within the radiation field are usually much longer than the scintillation probes, several centimeters versus a few millimeters at most for the scintillators \cite{Beaulieu-Review-2016}. Whether Cerenkov light requires removal in brachytherapy applications depends on the radioactive source used and the measurement geometry. In HDR brachytherapy with an $^{192}Ir$ source, Cerenkov radiation can cause large errors in dose reporting if it is not taken into account \cite{Therriault-2011, Therriault-Phantom-2011}. The production and removal of Cerenkov light in PSDs are widely discussed topics \cite{Beddar-water-equivalent-1992-1, Beddar-water-equivalent-1992-2, Boer-optical-1993, Fontbonne-2002, Clift-temporal-2002, Lambert-Cerenkov-2008, Archambault-MathForm-2012, Beaulieu-Review-2016}. In this study, the stem effect is accounted for.

Most studies that have characterized PSDs have been conducted using an optical fiber connected to a single point of measurement as the sensitive volume. Multiple scintillation sensors attached to a single optical chain have been used, but their application is limited to measurements made within 3 cm of an HDR brachytherapy source \cite{Archambault-MathForm-2012, Therriault-mPSD-2012, Patricia-2016}. New multipoint PSDs (mPSDs) could assess the dose at multiple points simultaneously, thereby improving treatment quality and accuracy. The multi-hyperspectral filtering method proposed by Archambault et al.\cite{Archambault-MathForm-2012} led to the conception of an mPSD in which each scintillator has an independent signal. Such an arrangement would allow simultaneous determination of the absorbed dose at different locations in a volume \cite{Therriault-mPSD-2012, Therriault-mPSD-Brachy-2013, Patricia-2016}.

The purpose of this study was to evaluate the performance of an mPSD in terms of sensitivity and accuracy, resulting in a thorough optimization of the optical chain, with applications to in vivo HDR brachytherapy in mind. To achieve this goal, three steps were followed. First, an experimental study was carried out to look for an optimal mPSD configuration. As the second step, we performed numerical optimizations to determine the proper configuration of the scintillation light detection system. In a third, and final stage, we evaluated the performance of the entire system in HDR brachytherapy.

\section{Materials and Methods}
%%\label{}

\subsection{Optical chain}

The optical chain in the proposed system has components that (1) generate scintillation light in the mPSD, (2) detect the scintillation light, and (3) analyze the signal. 

\begin{figure}
\centering
\begin{tabular}{c}
%trim={<left> <lower> <right> <upper>}
\includegraphics[trim = 1mm 1mm 1mm 0mm, clip, scale=0.38]{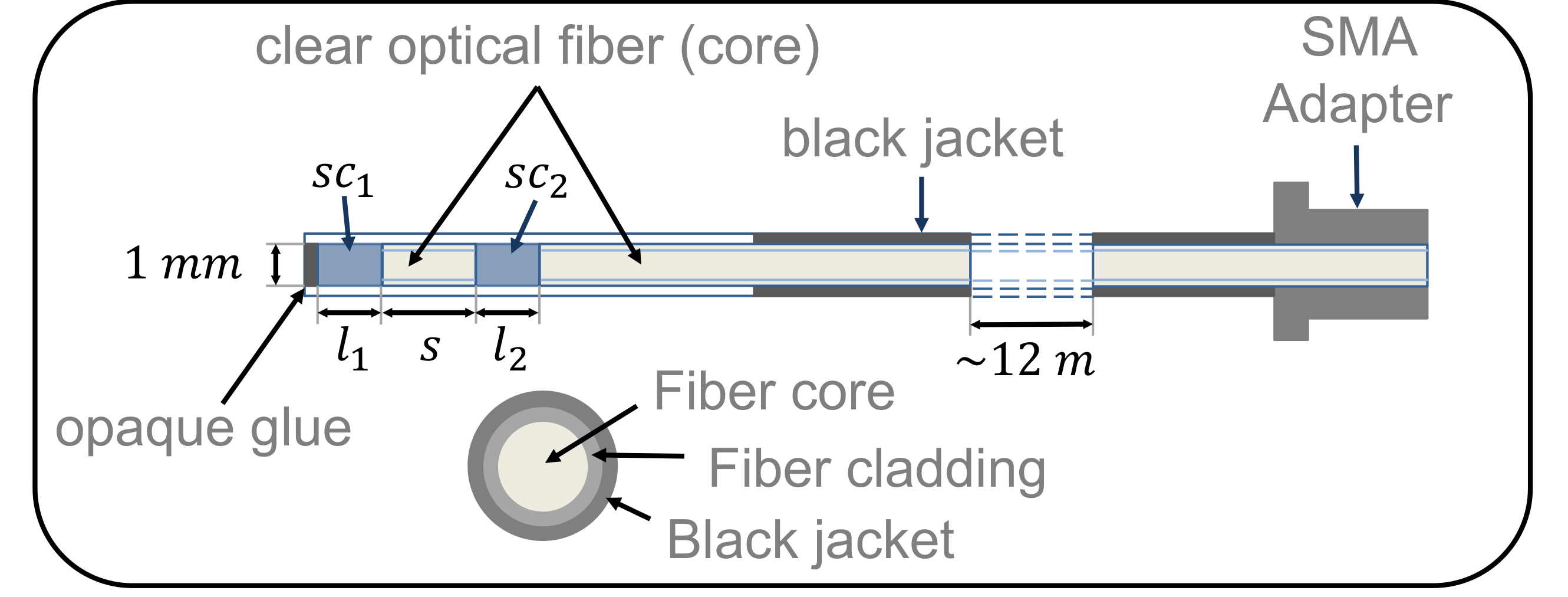}
\end{tabular}
\caption{\label{mPSD_det} Design schematic of an mPSD. $sc_1$ and $sc_2$ indicate scintillators 1 and 2, respectively; $l_1$ and $l_2$ denote the scintillators’ lengths. $s$ is the separation between the scintillators, 1 cm of clear optical fiber. Figure not to scale. SMA, subminiature version A.}
\end{figure}

The scintillating fibers used in this study were the plastic scintillators BCF-10, BCF-12, and BCF-60 from Saint Gobain Crystals (Hiram, OH, USA). Figure \ref{mPSD_det} shows the design schematic of a typical mPSD. The scintillators were separated from each other by 1 cm of clear optical fiber (Eska GH-4001, Mitsubishi Rayon Co., Ltd., Tokyo, Japan). The same type of fiber was also used to conduct the scintillating light to the photodetector surface. (Here, ``clear'' refers to a fiber in which no scintillation light is produced.) The aim of this study is to optimize a 3-point mPSD configuration. A single investigation with a 2-point mPSD was carried out, which is described in the following sections. All optical interfaces (scintillators and clear optical fibers) were polished using a SpecPro automated optical fiber polisher (Krell Technologies, Neptune City, NJ, USA) with successive grain sizes of 30 $\mu$m, 9 $\mu$m, 3 $\mu$m, and 0.3 $\mu$m. The detectors were constructed using a previously described coupling technique \cite{Ayotte-Surface-2006}.To ensure the reproducibility of the polishing and coupling techniques, we verified that the light collection across multiple detector samples does not vary more than 5 \%. Each 1-mm-diameter detector prototype was made light-tight using a black polyether block amide jacket from Vention Medical (Salem, NH, USA). The scintillating tip was sealed with a mixture of epoxy and black acrylic paint.

The scintillation light signal was guided to the photodetector surface by a clear optical fiber, which was attached to the photodetector with a subminiature version A connector (11040A, Thorslab, Newton, NJ, USA). Two types of photodetector were used: (1) an Ocean Optics QE65Pro spectrometer (Dunedin, FL, USA) and (2) a set of photomultiplier tubes (PMTs) coupled to dichroic mirrors and filters from Hamamatsu (Bridgewater, NJ, USA) \cite{PMT-Hamamatsu}. When the PMTs were used as the photodetector, the signal was read by a data acquisition board (NI USB-6289 M Series Multifunction I/O Device, National Instruments, Austin, TX, USA), and LabVIEW software version 15.0f2 (National Instruments) was used during the signal analysis stage \cite{DAQ-6289}.

\subsection{Experimental determination of the optimal mPSD configuration}

The purpose of this step was to identify the best configuration of scintillators in an mPSD. Interchangeable radiosensitive tips containing the scintillators were coupled to a clear optical fiber, which conveyed the light signal to the photodetector. We built and tested more than 20 mPSD prototypes using various combinations of scintillator positions and lengths to determine (1) the optimal position for each scintillator within the fiber and (2) the optimal length of each scintillator.

To evaluate the effect of scintillator position on the recorded signal, (A) the lengths of the scintillators inside the radiosensitive tips were kept identical (3 mm of BCF-10, 6 mm of BCF-12, 7 mm of BCF-60), and (B) the positions of the scintillators inside the fiber were altered. Each scintillator was placed in a distal, central, or proximal position relative to the photodetector surface. To determine the optimal scintillator length in terms of the scintillation light produced and spatial resolution, each scintillator was varied in length from 3 mm to 14 mm. The hyperspectral approach \cite{Archambault-MathForm-2012} dictated a balanced signal contribution from each scintillator to the overall light collection, while the spatial resolution of the dose required detection elements that were as small as possible. 

For each possible combination of scintillator number, position, and length, the total emission spectrum was obtained from simultaneous irradiation of all the scintillators in the fiber. Then, in order to obtain their individual contributions to the total emission spectrum, each scintillator was individually irradiated. Lead blocks were used to shield the neighboring scintillators from the incoming radiation. 

The spectral distributions were obtained with the spectrometer cooled to -20 $^\circ$C and the integration time set to 40 s. Background signals acquired prior to exposures were subtracted from the scintillation light signals.

The analysis in this section was performed under a fixed irradiation condition with an X-ray therapy system. A tube voltage of 120 kV (maximum photon energy of 120 keV) was selected to avoid Cerenkov light production. Several measurements were carried out in the low-energy range using an Xstrahl 200 (Xstrahl Ltd., Camberley, UK). Continuous beam irradiations with a tube current of 10 mA were performed according to the specifications shown in Table \ref{orthovoltage_unit}.

\begin{table}
\caption{\label{orthovoltage_unit} Main irradiation parameters when using a \textit{Xstrahl 200}  x-ray therapy system.}
\centering

\vspace{0.2cm}
	\small{\begin{tabular}{cccc}
	\hline
	\hline
	Energy      & $HVL_{1}$ & Added filtration           & Field size       \\
	   (kV)     &   (mm)    &      (mm)                  &   (cm)           \\
	\hline
	120         & 5.0 Al    & 0.5 Al$+$0.10 Cu           &   10$^\ast$       \\
	\hline
    \multicolumn{4}{c}{\small{\textit{HVL = Half-Value Layer}}} \\
	\multicolumn{4}{c}{$^\ast$ \small{\textit{diameter to circular shape}}} \\
	\end{tabular}}
\end{table}

\subsection{Determination of the optimal scintillation light detection system.}

In the second step, we sought the appropriate optical components for the light detection system used during measurements. Having a complete spectral characterization led us to perform a numerical analysis to determine the optical chain that would allow optimal scintillation light collection.

In our experimental set-up, scintillation light was read by an assembly of PMTs, which were coupled to a set of dichroic mirrors and filters that deconvolved the collected light into spectral bands. PMTs were chosen as the photodetectors because they have a high signal-to-noise ratio (SNR) and readout speed that overcomes many of the sensitivity issues of charge-coupled device-based systems. Generally, PMTs more accurately measure low light signals and have a faster response, making them more suitable for the demands of in vivo dosimetry applications \cite{Liu-real-time-2012, Boivin-photodetectors-2015, Boivin-2016}. Henceforth, an assembly composed of a dichroic mirror, filter, and PMT will be referred to as a ``channel''. From an optimization perspective, the signal produced in each channel was calculated, taking into account the measured scintillation spectrum and light yield, the manufacturer-reported transmission and attenuation of the optical components, and the experimentally characterized PMT noise. The experimental spectral characterization obtained for the mPSD constituted the main input. That spectral information was then used to construct the optical system and simulate its response when interacting with a radiation beam. No particle transport through Monte Carlo simulations was performed. A large set of possible component combinations (brute force) was explored to find the configuration that provided the best SNR. For the calculations, we used the characteristics of filters and dichroic mirrors from Hamamatsu series A10033 and A10034, respectively. For the PMTs, the models used corresponded to Hamamatsu series H10722. The numerical optimization took into account the fact that the number of channels depended on the number of scintillator points $N$ composing the mPSD and equaled $N + 1$. This procedure allowed us to optimize light transmission and to minimize the contribution of elements generating spurious light (as will be shown in Figure \ref{mPSD_sys_optimized}).

\subsection{Performance of the mPSD system in HDR brachytherapy}

With the optimized system, we next evaluated the performance of a 3-point mPSD in HDR brachytherapy. A Flexitron HDR afterloader (Elekta, Stockholm, Sweden) was used with an $^{192}Ir$ source. The cylindrical $^{192}Ir$ source pellet was 0.6 mm in diameter and 3.5 mm in length and was housed inside a stainless steel capsule 0.86 mm in diameter and 4.6 mm in length. The source air kerma strength was 32226 $U$. The HDR brachytherapy unit was remotely controlled and able to move the source to the desired position in a water tank by means of a 30 cm needle set from Best Medical International (Springfield, VA, USA). The mPSD dimensions allowed it to be inserted into an additional catheter for use during real-time dose verification.

\begin{figure}
\centering
\begin{tabular}{c}
%trim={<left> <lower> <right> <upper>}
\includegraphics[trim = 0mm 0mm 0mm 0mm, clip, scale=0.41]{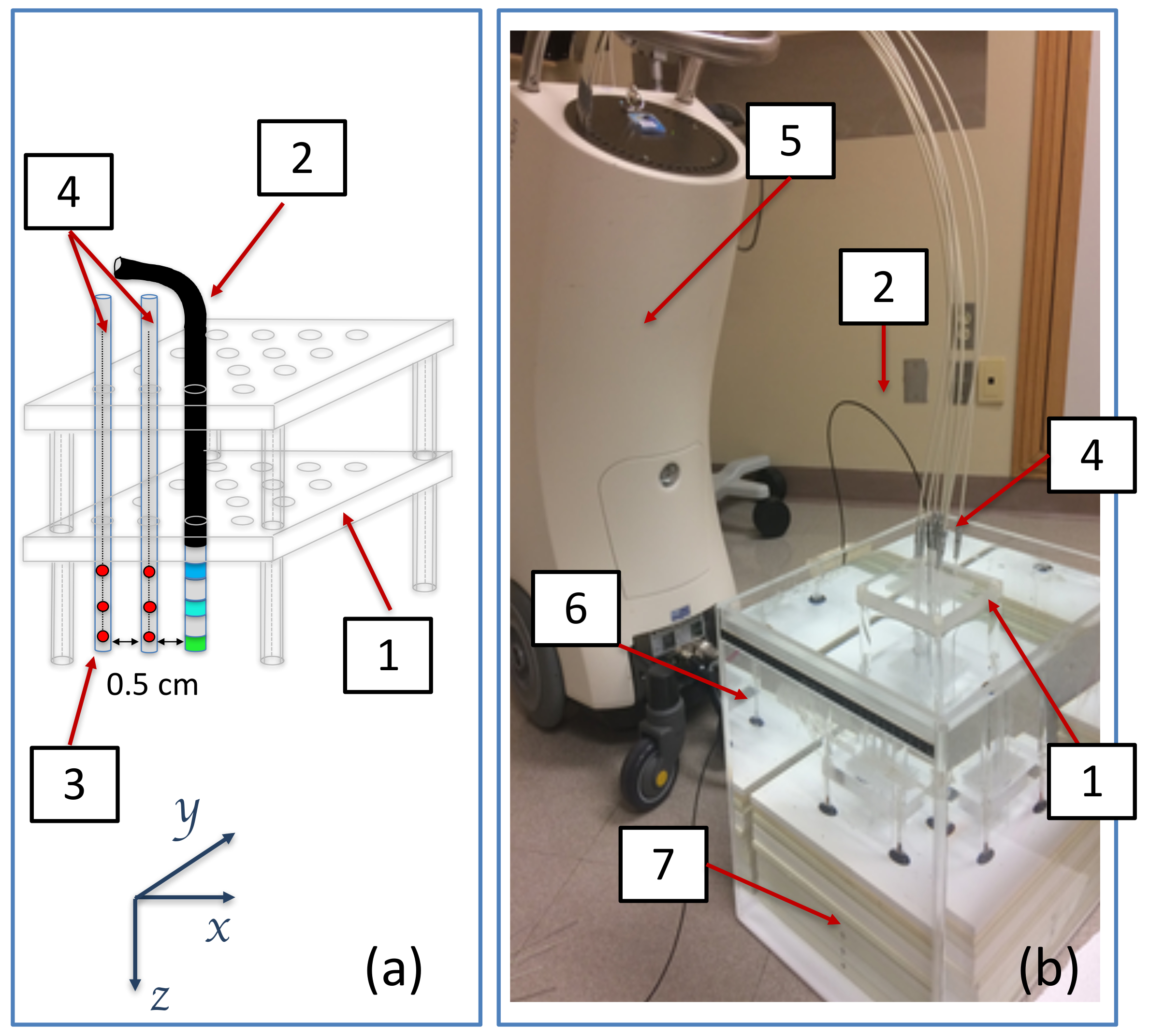}
\end{tabular}
\caption{\label{brachy_scheme} (a) Schematic of the poly(methyl methacrylate) (PMMA) phantom constructed for HDR brachytherapy measurements with an mPSD. The catheter positioning allowed source displacement parallel to the mPSD. (b) Experimental set-up for HDR brachytherapy measurements. (1) PMMA phantom, (2) mPSD, (3) $^{192}Ir$ source, (4) 30-cm catheters, (5) Flexitron HDR afterloader unit, (6) 40 $\times$ 40 $\times$ 40-cm$^3$ water tank, (7) solid-water slabs.}
\end{figure}

To be consistent with the TG-43U1 formalism \cite{TG-43-Update}, measurements were performed with the source and detector isotropically covered by at least 20 cm of water. To ensure the accuracy and reproducibility of the source-to-detector distance, all the catheters were inserted in a custom-made poly(methyl methacrylate) phantom (Figure \ref{brachy_scheme}a), which was in turn placed inside a 40 $\times$ 40 $\times$ 40-cm$^3$ water tank. As shown in Figure \ref{brachy_scheme},the phantom was composed of 2 catheter insertion templates of 12 $\times$ 12 cm$^2$, separated by 20 cm. This phantom allowed for source-to-detector parallel displacement. Figure \ref{brachy_scheme}b shows the experimental set-up used during measurements. 10 catheters were inserted in the template phantom at distances ranging from 0.5 cm to 7.0 cm away from the mPSD catheter. Three dwell positions per catheter were planned.  Following the axis convention shown in figure \ref{brachy_scheme}a, the source $z$ locations were chosen relative to the effective center of each scintillator volume.

The dosimetric system was initially calibrated under the same conditions used to perform the measurements, following the TG-43U1 \cite{TG-43-Update} and hyperspectral \cite{Archambault-MathForm-2012} formalisms. The luminescence dosimetry system was controlled using LabVIEW software. A gain input voltage of 1 V was assigned to each PMT, producing a channel output of 2$\times$10$^6$. The linear relationship between the input voltage and the gain was assessed for voltage between 0.5 V and 1.1 V. For each measurement channel, 70000 samples per second were acquired. Dose values were recorded in real time by the mPSD. All measurements were repeated at least 5 times, and the measurement set-up was completely unmounted between each measurement. Statistical variations in the readings were determined by setting a source dwell time of 60 s per dwell position.

\begin{figure}
\centering
\begin{tabular}{c}
%trim={<left> <lower> <right> <upper>}
\includegraphics[trim = 0mm 0mm 0mm 0mm, clip, scale=0.35]{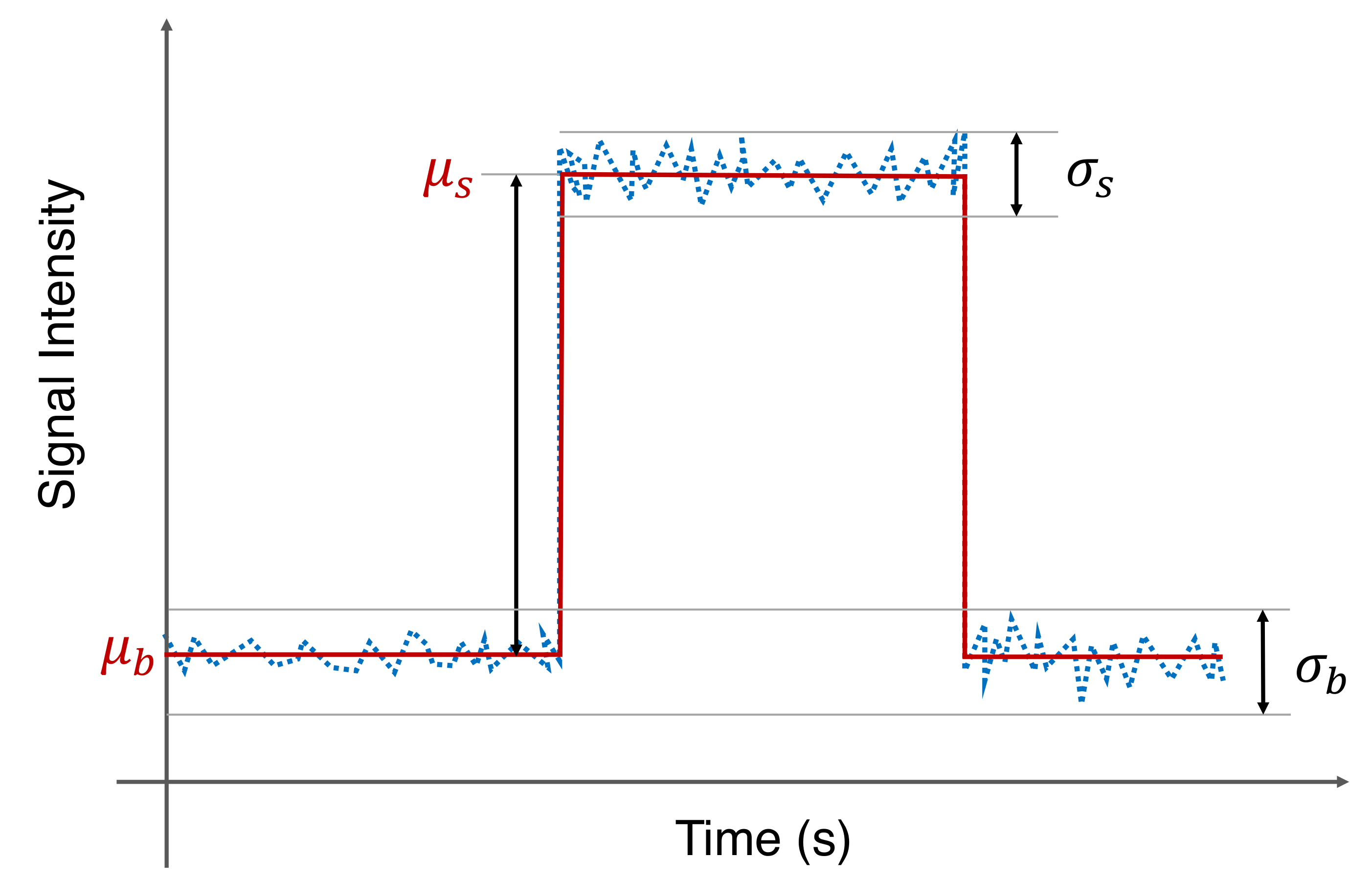} \\
\end{tabular}
\caption{\label{scheme_pulse} Typical signal pulse used for signal-to-noise ratio and signal-to-background ratio calculations. The indicated values are: $\mu_{s}$, mean signal value; $\mu_{b}$, mean background value; $\sigma_{s}$, signal standard deviation; $\sigma_{b}$, background standard deviation and the difference between $\mu_{s}$ and $\mu_{b}$.}
\end{figure}

The sensitivity of the dosimetry system was evaluated using the SNR and the signal-to-background ratio (SBR) associated with each scintillator during HDR brachytherapy measurements. Figure \ref{scheme_pulse} is a representation of a typical signal pulse, showing the magnitudes that were used in SNR and SBR determination: $\mu_{s}$, the mean signal; $\mu_{b}$, the mean background signal;  $\sigma_{s}$ the signal standard deviation; and  $\sigma_{b}$ the background standard deviation.

\begin{equation}
\label{SNR_eq}
	SNR = \frac{\mu_{s}}{\sigma_{s}}
\end{equation}
\begin{equation}
\label{SBR_eq}
	SBR = \frac{\mu_{s}}{\mu_{b}}
\end{equation}

The SNR is a commonly used metric for characterizing the global performance of optoelectronic systems. In the case of PSD performance assessment, the noise term includes Cerenkov radiation. A few SNR studies using PMTs as the photodetector have been performed \cite{Beddar-light-output-2003, Lacroix-SNR-2009, Boivin-photodetectors-2015}. The scintillator SNR as function of dose rate was obtained using equation \ref{SNR_eq}, where the numerator represents the mean signal for a determined irradiation in a fixed time, and the denominator is the standard deviation of the collected signal. SBR was determined according to equation \ref{SBR_eq} and is the ratio of the mean signal value  and the mean background value  for a fixed irradiation time.

\subsection{Cerenkov radiation removal}

 In our study, multiple probes were read by a single clear collecting optical fiber; thus, we used the hyperspectral filtering technique proposed by Archambault et al.  \cite{Archambault-MathForm-2012} as the stem effect removal method during HDR brachytherapy measurements. Equations \ref{raw_m}, \ref{D_system} and \ref{Dose_multi} represent the removal method for an mPSD configuration. Dose calculation in mPSDs is based on the assumption that the recorded signal results from the linear superposition of spectra; no self-absorption interactions among the scintillators composing the mPSD are considered \cite{Archambault-MathForm-2012}. The idea behind this formalism is that once the light emission of each component at different wavebands is known, the total signal recorded can be decoupled, and the signal fraction contributed by each scintillator can be determined. 

\begin{equation}
\label{raw_m}
	m = Rx
\end{equation}

\begin{equation}
\label{D_system}
\left(
\begin{array}{cccc}
d_{p1,1} & d_{p1,2} & d_{p1,3} & d_{p1,4} \\
d_{p2,1} & d_{p2,2} & d_{p2,3} & d_{p2,4} \\
d_{p3,1} & d_{p3,2} & d_{p3,3} & d_{p3,4} \\
\end{array}
\right) =
\left(
\begin{array}{cccc}
C_{p1,L_{1}} & C_{p1,L_{2}} & C_{p1,L_{3}} & C_{p1,L_{4}} \\
C_{p2,L_{1}} & C_{p2,L_{2}} & C_{p2,L_{3}} & C_{p2,L_{4}} \\
C_{p3,L_{1}} & C_{p3,L_{2}} & C_{p3,L_{3}} & C_{p3,L_{4}} \\
\end{array}
\right) \times
\left(
\begin{array}{cccc}
m_{L_{1},1} & m_{L_{1},2} & m_{L_{1},3} & m_{L_{1},4} \\
m_{L_{2},1} & m_{L_{2},2} & m_{L_{2},3} & m_{L_{2},4} \\
m_{L_{3},1} & m_{L_{3},2} & m_{L_{3},3} & m_{L_{3},4} \\
m_{L_{4},1} & m_{L_{4},2} & m_{L_{4},3} & m_{L_{4},4} \\
\end{array}
\right)
\end{equation}

\begin{equation}
\label{Dose_multi}
	d' = C \times m'
\end{equation}

A raw measurement $m$ with the mPSD system  is a function of a given photon flux $x$ (the number of photons emitted for a given emission source, either scintillating elements or any other source of light) and the system response matrix $R$. In equation \ref{raw_m}, $m$ is a vector of $L$ elements; $R$ is $L \times N$ dimensions, and $x$ is a vector of $N$ elements. $L$ represents different wavelength filters or channels. The number of measurement channels $L$ should be equal to $N + 1$. The additional channel is included to take into account the stem effect, which should be removed from the measured signal \cite{Therriault-2011}.

Equation \ref{D_system} is the mathematical equation for determination of the calibration factor. The dose $d_{i;k}$ received by the scintillator during irradiation is directly proportional to the number of scintillation photons in the absence of losses (quenching); for this reason, $d_{i,k} = a_i x_{i,k}$, $a_i$ being a proportionality constant and $x_{i,k}$ the photon fluence in the scintillating material $i$ during the measurement at position $k$. However, knowing the dose at a specific point requires a previous calibration to determine the calibration factor $C$ for each scintillation point as well as each measurement channel. For such a calibration the dose (e.g., $d_{p1, p2, p3, p4}$ ) should be known at each point \textit{p}. We calculated these dose values by using the TG-43U1 formalism \cite{TG-43-Update}. To account for the finite size of each scintillator, TG-43U1 dose values were integrated over each scintillator's sensitive volume. 

Once the calibration factor $C$ is known, the dose $d'$ at each point can be determined using equation \ref{Dose_multi}, where $m'$ represents the raw data acquired during measurements. The apostrophe in equation \ref{Dose_multi} is used to highlight that this is a new set of measurements, the goal of which is to determine the absorbed dose, not the calibration factor, which is already known at this stage.

\section{Results and discussion}

\subsection{mPSD optimal configuration}
\subsubsection{Scintillator position within the fiber}

\begin{figure}
\centering
%trim={<left> <lower> <right> <upper>}
\includegraphics[trim = 0mm 0mm 0mm 0mm, clip, scale=0.45]{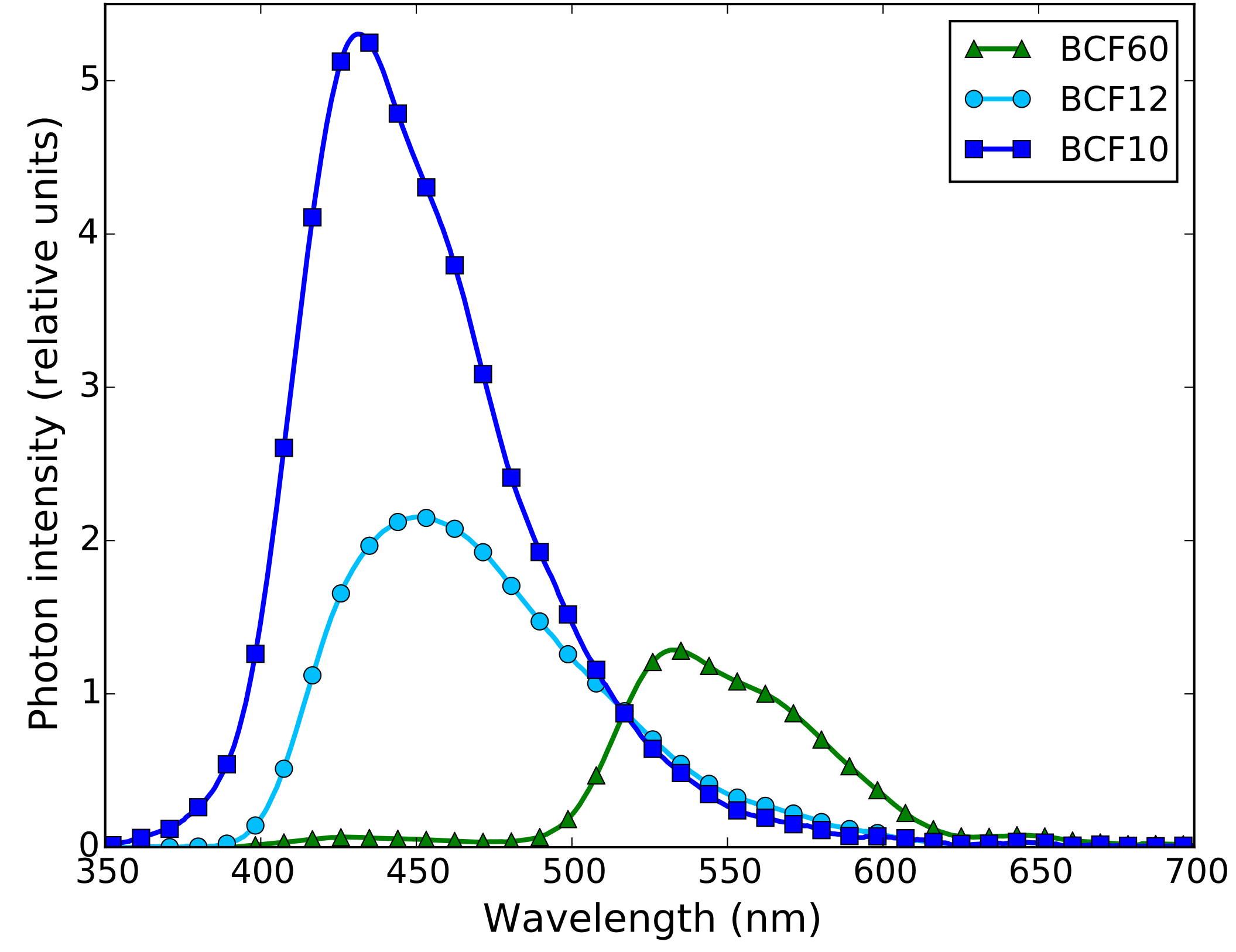}
\caption{\label{scint_spectra} Individual scintillator spectra normalized to a 1-mm-long scintillator.}
\end{figure}

A single investigation was performed with a 2-point mPSD configuration to determine the optimal order of scintillators in the mPSD. Each scintillator spectrum was measured independently. Figure \ref{scint_spectra} shows the individual spectra with intensities normalized to a 1-mm scintillator length. As shown in the figure, the scintillation intensity was strongest in the BCF-10 scintillator, whereas the BCF-60 scintillator had the weakest scintillation intensity.

In all the tested combinations, the scintillator's individual spectra evidenced no self-absorption or cross-excitation effects. The main differences observed related to the position occupied by each scintillator inside the fiber. Table \ref{2p_signal_proportion} shows the measured signal proportion for different 2-point mPSD configurations. For each configuration, the proportion of the total signal coming from each independent scintillator was calculated by determining the area under the curve. The signal proportion was more balanced when a BCF-60 scintillator was placed in the distal position and coupled to a BCF-10 scintillator. At 530 nm, the BCF-60 signal represented 37\% of the total signal when coupled to a BCF-10 scintillator but only 9\% when coupled to a BCF-12 scintillator. Therefore, this combination of scintillators and positions is recommended for a 2-point mPSD.

\begin{table}
\caption{\label{2p_signal_proportion} Scintillator signal proportion (in \%) for each combination of scintillators and positions inside the optical fiber.}
\centering

\vspace{0.2cm}
	\begin{tabular}{ccc|ccccc}
	\hline
	\hline
	&  &  & \multicolumn{5}{c}{\textbf{Proximal}} \\
	&  &  & BCF-10  & & BCF-12  & & BCF-60   \\
	\hline 
	\multirow{3}{0.1cm}{\rotatebox{90}{\textbf{Distal}}} & & BCF-10 &  - & &  8.6/91.4 & & 17.6/82.4 \\
	& & BCF-12 & 18.8/81.2 & & - & & 7.2/93.8  \\
	& & BCF-60 & 37.4/62.6 & & 8.8/91.2 & & -  \\
	\hline
	\end{tabular}
\end{table}

\begin{figure}
\centering
\begin{tabular}{cc}
%trim={<left> <lower> <right> <upper>}
\includegraphics[trim = 0mm 0mm 0mm 0mm, clip, scale=0.55]{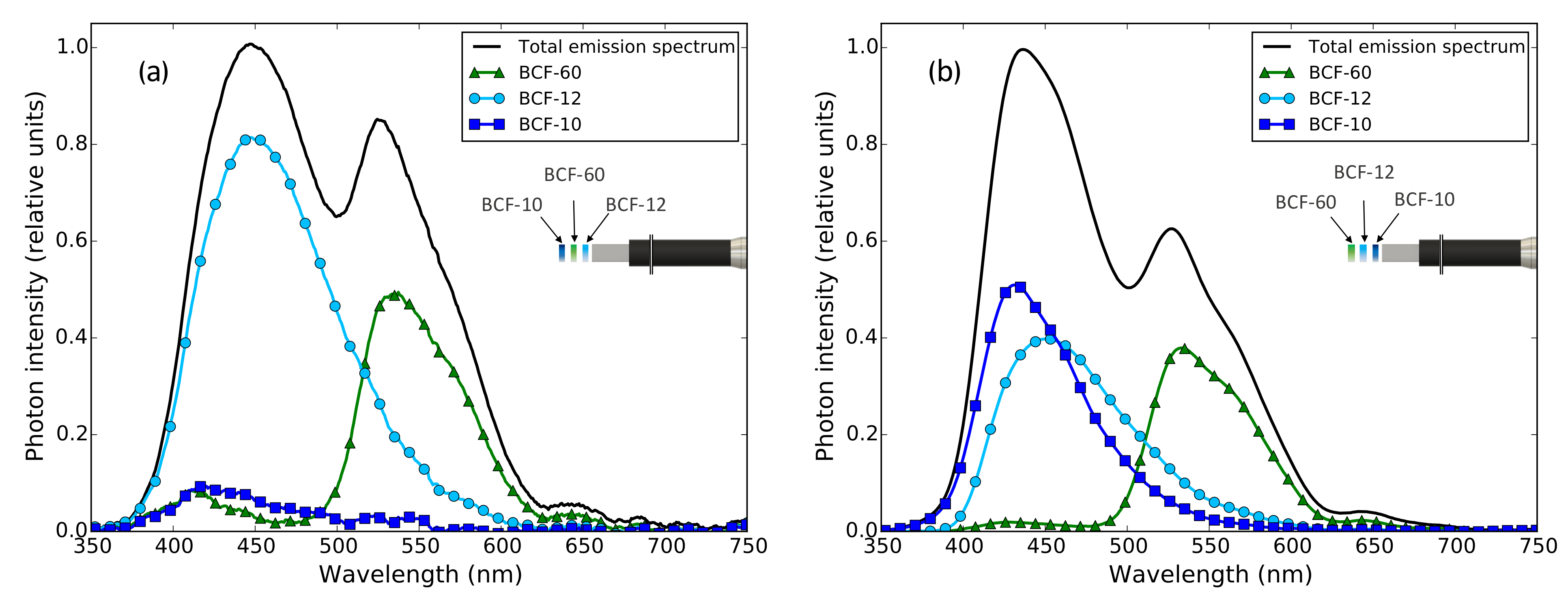}
\end{tabular}
\caption{\label{Stokes_shift} (a) Stokes shift effect. (b) Spectral distributions obtained with all scintillators in the optimal positions.}
\end{figure}

The signal analysis demonstrated that the shorter wavelength scintillator should always be placed closer to the photodetector and the longer wavelength scintillator in the distal position. Because of the Stokes shift, the absorption spectrum always has a lower wavelength range than the emission spectrum. If the aforementioned configuration is not used, inter-scintillator excitation and self-absorption effects can take place, and as a consequence, the light transmission through the collecting fiber is not optimal. To exemplify this effect, 3-point mPSDs with 2 different configurations of scintillator positions inside the fiber were constructed. Their spectral distributions are shown in Figure \ref{Stokes_shift}. Figure \ref{Stokes_shift}a shows the spectra with the BCF-10 placed at the distal position, the BCF-60 in the center, and the BCF-12 in the proximal position. Almost all of the light produced by the BCF-10 scintillator was absorbed by the neighboring scintillators, whose photon intensities were higher than they were in the optimal configuration (Figure \ref{Stokes_shift}b). Hence, in the subsequent experiments, we used mPSDs in which the scintillators were placed inside the optical fiber in decreasing order of wavelength from distal to proximal positions.

\begin{figure}
\centering
%trim={<left> <lower> <right> <upper>}
\includegraphics[trim = 0mm 0mm 0mm 0mm, clip, scale=0.4]{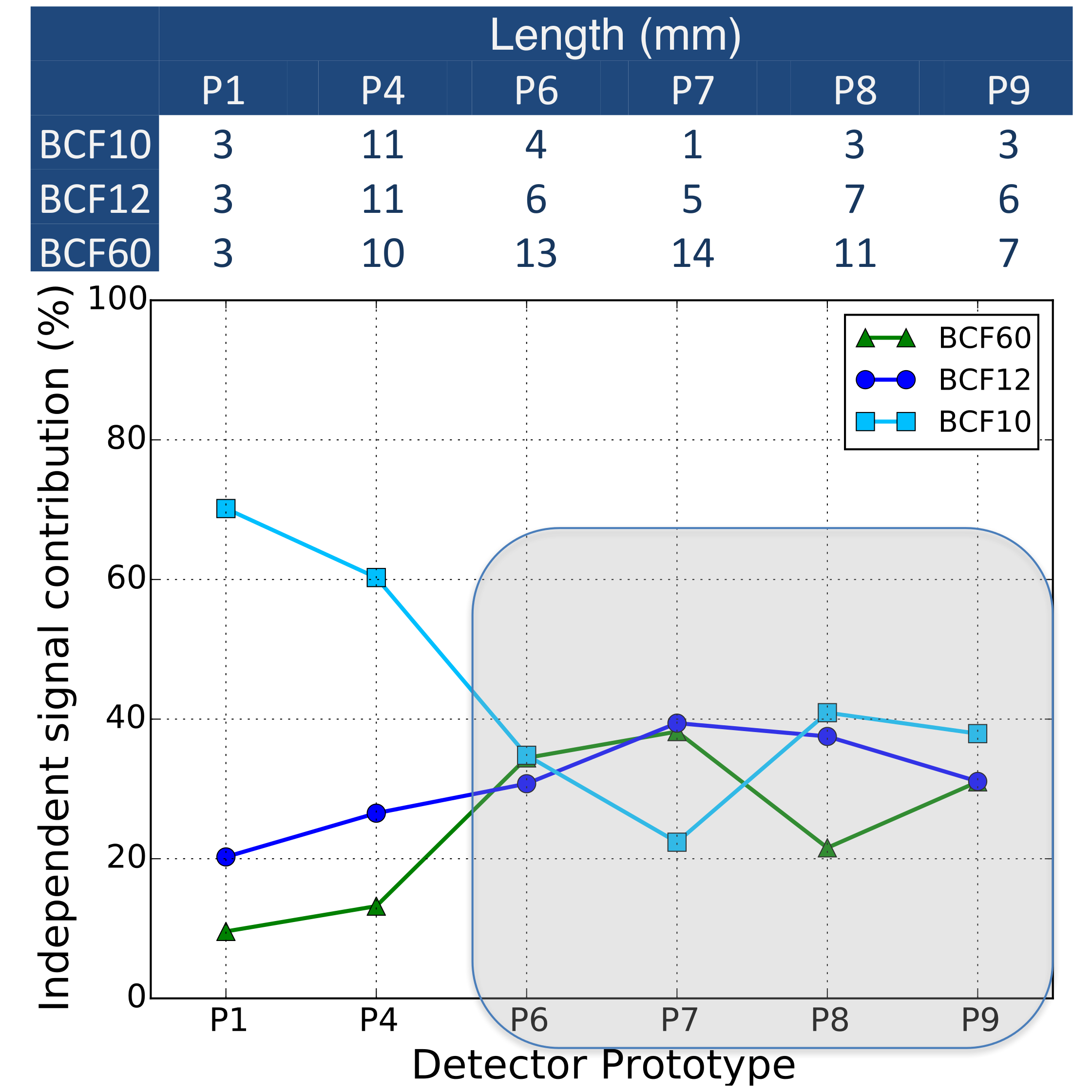}
\caption{\label{3p_width_effect} Fraction of the total scintillation light produced by each scintillator as a function of its length. The shaded region indicates the combination of sensor lengths that result in balanced signals.}
\end{figure}

In the case of a 3-point mPSD composed of 3 mm of BCF-60 at the distal position, BCF-12 in the center, and BCF-10 at the proximal position (mPSD prototype P1 in Figure \ref{3p_width_effect}), we observed no self-absorption or cross-excitation effects, but the scintillator's independent signals were not balanced at all. In such a case, the scintillation process is more efficient in BCF-10 than in the other scintillators, accounting for almost 71\% of the total signal. The intensities of BCF-12 and BCF-60 were closer to one another, with 20\% and 9\% of the total signal, respectively.

\subsubsection{Optimal scintillator length}

The intensity of the measured scintillation light depends on the scintillator size, the coupling method, the fiber core size, and the fiber numerical aperture. To determine the optimal length of each scintillator, 9 different 3-point mPSD prototypes were constructed. Figure \ref{3p_width_effect} shows the contributions of individual scintillator signals for each of the 3-point mPSDs and specifies the length of each scintillator in millimeters. As indicated by the shadowed region in Figure \ref{3p_width_effect}, detector configurations P6 to P9 provided the required balanced signals for optimal hyperspectral deconvolution. P9 was selected as the optimal detector because it also minimized variations in sensor length. In the P9 mPSD, the BCF-10 scintillator was 3 mm long, the BCF-12 scintillator was 6 mm long, and the BCF-60 scintillator was 7 mm long.

\subsection{Optimized scintillation light detection system}

\begin{figure}
\centering
\begin{tabular}{c}
%trim={<left> <lower> <right> <upper>}
\includegraphics[trim = 0mm 0mm 0mm 0mm, clip, scale=1.5]{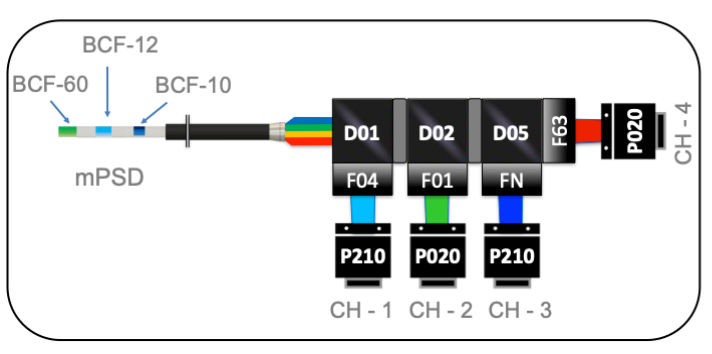}
\end{tabular}
\caption{\label{mPSD_sys_optimized} Schematic of the light collection system obtained from calculations. All the components used were from Hamamatsu. D indicates dichroic mirrors from series A10034, F indicates filters from series A10033, and P210 and P020 refer to PMTs H10722 210 and 020, respectively. CH indicates the channel number. FN refers to a filter that transmits 100\% of the incoming light to the photodetector in the wavelength range of 300 nm to 500 nm}.
\end{figure}

Following the determination of the optimal length, the numerical optimization allowed us to determine the best combination of components to be used for the measurements of the light collection system. Figure \ref{mPSD_sys_optimized} shows a schematic of the appropriate arrangement of the components of the light collection system obtained from these calculations. 

\begin{figure}
\centering
%trim={<left> <lower> <right> <upper>}
\includegraphics[trim = 0mm 0mm 0mm 0mm, clip, scale = 0.41]{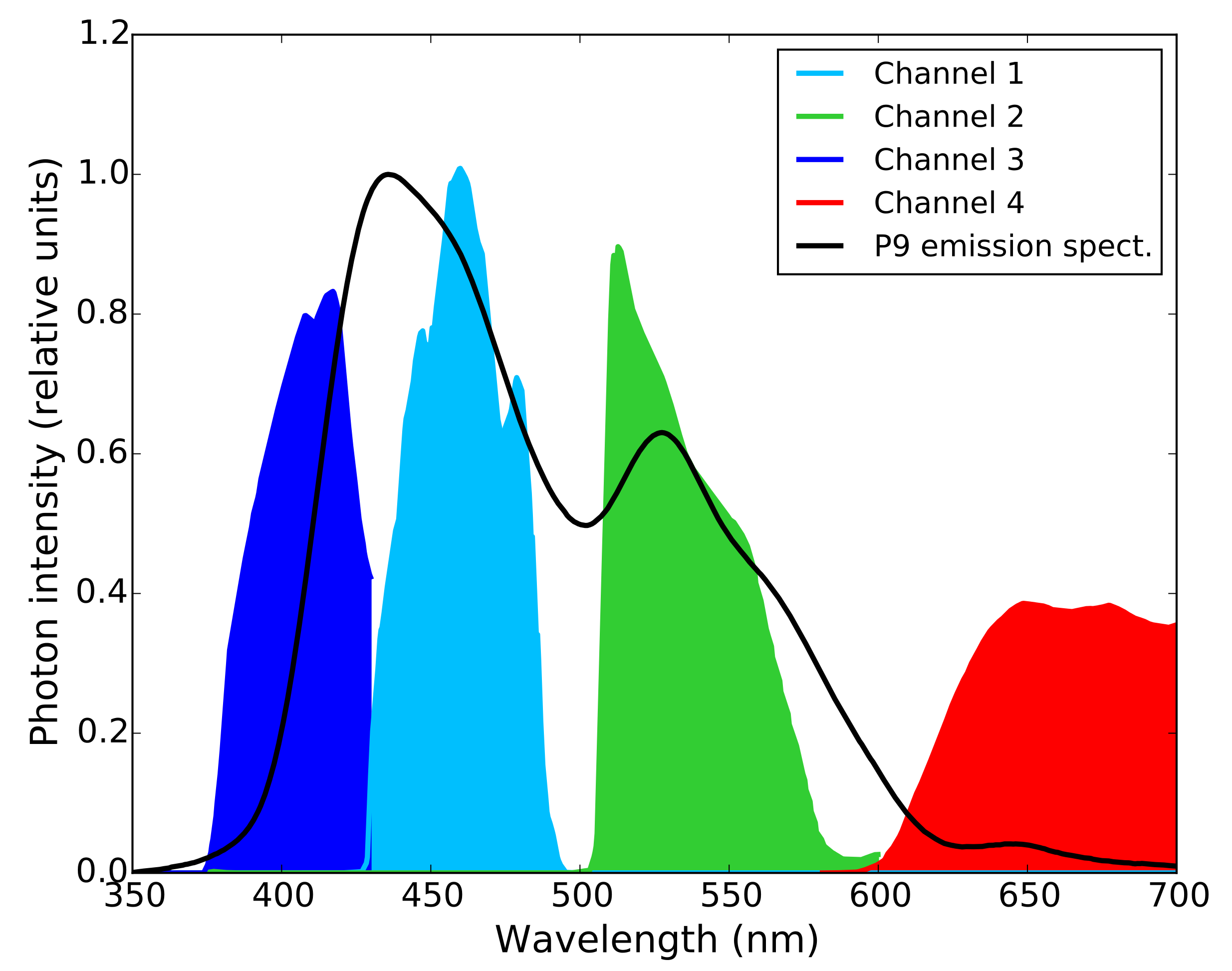}
\caption{\label{3p_mod_channel} Emission and filtered spectra produced by the P9 mPSD prototype used for absorbed dose determination. The filtered spectra for each channel were obtained by applying optical filtration to the light entering each PMT.The intensities in the emission and filtered spectra are normalized to their maximum intensity.}
\end{figure}

This assembly filtered the total emission spectrum from the mPSD to produce a filtered spectrum entering each PMT. The PMT's voltage output is then used to calculate the absorbed dose in the absence of Cerenkov radiation. Figure \ref{3p_mod_channel} shows each channel's filtered spectrum and the total emission spectrum of the scintillation light generated by the P9 mPSD.

The study done by Therriault-Proulx et al. \cite{Therriault-mPSD-Brachy-2013} in HDR brachytherapy uses the same scintillating elements described in this work (BCF-10, BCF-12, and BCF-60; Saint-Gobain Crystals, Hiram, OH), but with different positions inside the optical assembly. In that work it was suggested to use the BCF-10 in the central position while BCF -12 at the proximal position with respect to the photodetector surface. Based on the signal analysis performed in this study, we propose an optimized mPSD design that maximizes the scintillation light collection, resulting in the configuration shown in  Figure \ref{mPSD_sys_optimized}, which inverts the BCF-10 and BCF-12 positions.

\subsection{Performance of the mPSD system in HDR brachytherapy}

\subsubsection{Absorbed dose as function of distance}

Dose distributions in terms of distance to an HDR brachytherapy source were obtained, with the P9 mPSD calibrated at 1.5 cm from the $^{192}Ir$ source. This calibration distance represented a compromise between measurement uncertainties and positioning uncertainties. Andersen et al. \cite{Andersen-time-resolved-2009}  demonstrated that positioning uncertainty dominates in measurements close to the source, whereas measurement uncertainty dominates at long distances. In order to ensure that we had enough data in the response recording, a source dwell time of 60 s was used at each dwell position. 

\begin{figure}
\centering
%trim={<left> <lower> <right> <upper>}
\includegraphics[trim = 0mm 0mm 0mm 0mm, clip, scale = 0.45]{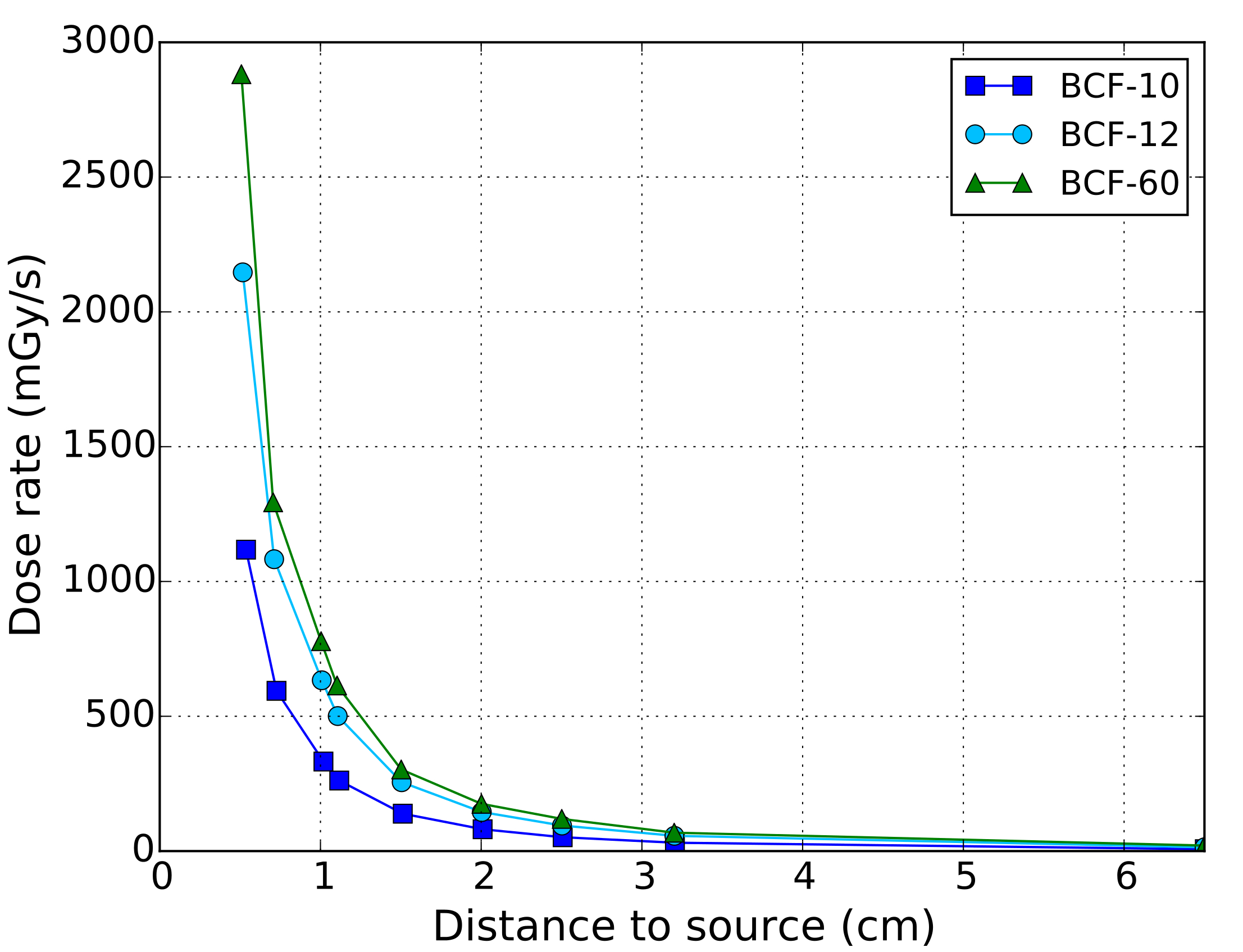} 
\caption{\label{mPSD_brachy_dose} Scintillator dose rates as a function of distance to the HDR brachytherapy source. The distances to the source are relative to each scintillator effective center.}
\end{figure}

Figure \ref{mPSD_brachy_dose} shows dose rate readings for each scintillator, and Table \ref{mPSD_brachy_Dev} details the standard deviations for each distance to the brachytherapy source. For all 3 scintillators, the standard deviations were generally no greater than 5\% of the mean dose reading, although this value, as expected, increased with distance from the source. At a distance of 6.5 cm, the standard deviation exceeded 10\% for all scintillators. At that distance, the source radiation does not produce enough scintillation in the mPSD, so the recorded signal can be considered to be background. Nonetheless, the absolute standard deviation was small relative to the mean dose.

\begin{table}
\caption{\label{mPSD_brachy_Dev} Standard deviation (SD) of 3-point mPSD measurements and deviation of the mean measured dose from the values predicted by TG-43U1. The distances to the source are relative to each scintillator effective center.}
\centering

\vspace{0.2cm}
	\begin{tabular}{ccccc}
	\hline
	\hline
	& \textbf{Distance} & \textbf{Dose per}   &                          & \textbf{Deviation}  \\
	& \textbf{to source}& \textbf{1s reading} & \textbf{SD}              & \textbf{from TG-43U1}  \\
	& (cm)              & (mGy)               & \hspace{0.1cm}(\%) \hspace{0.1cm} (mGy)             &  \hspace{0.1cm}(\%) \hspace{0.1cm} (mGy)      \\
	\hline 
	\multirow{8}{0.1cm}{\rotatebox{90}{\textbf{BCF-10}}} &	0.5	&	1117.9	&	0.7	\hspace{0.3cm}	7.7	&	3.0	\hspace{0.3cm}	33.2	\\	
	&	0.7	&	594.4	&	0.7	\hspace{0.3cm}	4.2	&	1.1	\hspace{0.3cm}	6.5	\\
	&	1.0	&	332.0	&	1.2	\hspace{0.3cm}	4.1	&	1.1	\hspace{0.3cm}	3.6	\\
	&	1.1	&	261.6	&	1.6	\hspace{0.3cm}	4.2	&	3.7	\hspace{0.3cm}	9.7	\\
	&	1.5	&	138.7	&	2.8	\hspace{0.3cm}	3.9	&	3.8	\hspace{0.3cm}	5.3	\\
	&	2.0	&	80.9	&	4.2	\hspace{0.3cm}	3.4	&	5.5	\hspace{0.3cm}	4.5	\\
	&	2.5	&	51.3	&	4.5	\hspace{0.3cm}	2.3	&	7.5	\hspace{0.3cm}	3.9	\\
	&	3.2	&	31.0	&	4.1	\hspace{0.3cm}	1.3	&	7.3	\hspace{0.3cm}	2.3	\\
	&	6.5	&	6.0	&	12.8 \hspace{0.3cm} 	0.8	&	21.3 \hspace{0.3cm} 	1.3	\\
	&           &           &                   &           \\
	\multirow{8}{0.1cm}{\rotatebox{90}{\textbf{BCF-12}}} &	0.5	&	2146.6	&	0.5	\hspace{0.3cm}	10.3	&	1.1	\hspace{0.3cm}	23.2	\\	
	&	0.7	&	1082.6	&	0.7	\hspace{0.3cm}	7.2	&	0.8	\hspace{0.3cm}	9.1	\\
	&	1.0	&	633.6	&	0.8	\hspace{0.3cm}	5.3	&	2.5	\hspace{0.3cm}	15.6	\\
	&	1.1	&	500.9	&	0.8	\hspace{0.3cm}	3.9	&	2.2	\hspace{0.3cm}	11.1	\\
	&	1.5	&	255.4	&	2.0	\hspace{0.3cm}	5.2	&	2.0	\hspace{0.3cm}	5.1	\\
	&	2.0	&	144.8	&	4.2	\hspace{0.3cm}	6.1	&	2.2	\hspace{0.3cm}	3.2	\\
	&	2.5	&	94.6	&	3.1	\hspace{0.3cm}	2.9	&	2.0	\hspace{0.3cm}	1.9	\\
	&	3.2	&	56.6	&	3.2	\hspace{0.3cm}	1.8	&	2.3	\hspace{0.3cm}	1.3	\\
	&	6.5	&	13.3	&	11.2 \hspace{0.3cm} 	1.5	&	12.4 \hspace{0.3cm} 	1.6	\\
	&           &           &                   &           \\
	\multirow{8}{0.1cm}{\rotatebox{90}{\textbf{BCF-60}}} &	0.5	&	2881.6	&	0.4	\hspace{0.3cm}	11.3	&	2.3	\hspace{0.3cm}	66.9	\\	
	&	0.7	&	1293.6	&	0.7	\hspace{0.3cm}	8.5	&	0.6	\hspace{0.3cm}	7.2	\\
	&	1.0	&	778.5	&	1.2	\hspace{0.3cm}	9.4	&	3.4	\hspace{0.3cm}	26.7	\\
	&	1.1	&	614.3	&	1.2	\hspace{0.3cm}	7.1	&	4.3	\hspace{0.3cm}	26.3	\\
	&	1.5	&	302.6	&	1.7	\hspace{0.3cm}	5.0	&	0.8	\hspace{0.3cm}	2.5	\\
	&	2.0	&	175.2	&	6.5	\hspace{0.3cm}	11.4	&	1.0	\hspace{0.3cm}	1.8	\\
	&	2.5	&	119.2	&	5.0	\hspace{0.3cm}	5.9	&	2.4	\hspace{0.3cm}	2.9	\\
	&	3.2	&	68.1	&	6.8	\hspace{0.3cm}	4.6	&	1.7	\hspace{0.3cm}	1.2	\\
	&	6.5	&	21.9	&	20.9 \hspace{0.3cm} 	4.6	&	13.7 \hspace{0.3cm} 	3.0	\\
	\hline
	\end{tabular}
\end{table}

TG-43U1 dose values were used as a reference; the last column in Table \ref{mPSD_brachy_Dev} presents the differences between the measured dose and the TG-43U1 dose values at each distance to the brachytherapy source. In general, the measured mPSD dose and the TG-43U1 dose agreed well at short distances to the source, but the difference increased as the source moved away from the mPSD.

\subsubsection{Evaluation of scintillation signal and system sensitivity}

\begin{figure}
\centering
\includegraphics[trim = 0mm 0mm 0mm 0mm, clip, scale = 0.43]{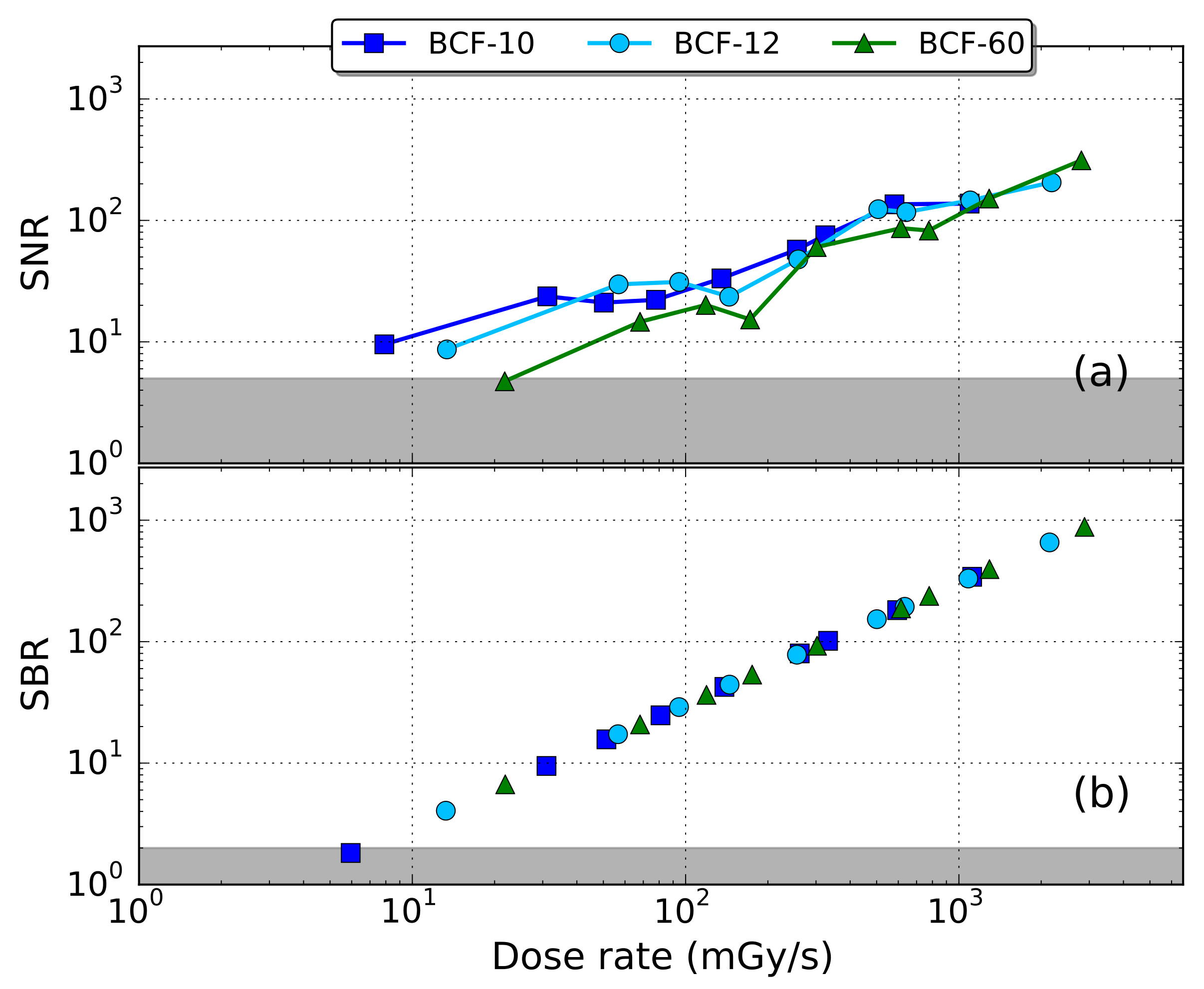} 
\caption{\label{SNR_SBR_dose} Signal-to-noise ratio (SNR) and (b) signal-to-background ratio (SBR) as a function of dose rate for BCF-10, BCF-12, and BCF-60 scintillators in the P9 mPSD prototype. Shaded areas at the bottom of the graphs indicate the cutoff values for each parameter.}
\end{figure}

Figure \ref{SNR_SBR_dose}a shows the SNR in terms of dose rate for each scintillator in detector P9. According to the Rose criteria, proper recognition (detection) of an object strongly depends on SNR, only becoming possible when SNR exceeds 5; detection performance degrades as SNR approaches zero \cite{Bushberg-3rd-Ed}. Thus, an SNR of 5 was the minimum sensitivity considered in this study. The BCF-10 and BCF-12 scintillators produced an SNR greater than 5 at all distances to the source. At dose rates below 22 mGy/s, the SNR produced by the BCF-60 scintillator fell below 5. These data suggest that, with regard to SNR, the dosimetric system characterized in this study is sensitive enough to measure dose rates above 22 mGy/s at distances to the source below 6.4 cm.

SNR analysis of PSD responses for various photodetectors was conducted by Boivin et al. \cite{Boivin-photodetectors-2015} in the energy range of clinical interest. SNR values in a range of 100 to 1000 for dose rates between 0.1 mGy/s and 30 mGy/s were reported, being around 2 orders of magnitude greater than the results obtained in the low dose rate range of this study. The differences are easily explained by the differences in design. First, the study done by Boivin et al.\cite{Boivin-photodetectors-2015} was conducted with a single point configuration detector that only had a single coupling interface, while the one reported in this study has 5 coupling interfaces intrinsicly decreasing the overall light collection in mPSD. Secondly, the scintillator size used in the study done by Boivin et al. \cite{Boivin-photodetectors-2015} was 10 mm vs. a maximum size of 7 mm in the mPSD. Thirdly and most important, the system used here is subject to optical filtration of the light produced; Boivin et al. \cite{Boivin-photodetectors-2015} used a system with no optical filtration, recording the signal of a PSD directly on a PMT module. 

To evaluate how well the dosimetric system differentiated a signal pulse from the background signal, SBR values were calculated at each dose rate. Furthermore, several signal acquisitions were performed without irradiation  resulting in a background signal value of $\mu_b$ $\pm$ 0.16 \%. To properly differentiate signal from background, a minimum SBR of 2 is required. Figure \ref{SNR_SBR_dose}b shows the SBR results obtained for each scintillator in the mPSD. The SBR for the BCF-10 scintillator fell below the SBR cutoff value for dose rates of around 6 mGy/s at 6.5 cm relative to the source, with an SBR of 1.8. The SBR is directly proportional to the photon fluence from scintillation, which is in turn proportional to the scintillator volume. According to the previously determined optimal mPSD design, the length of the BCF-10 scintillator was only 3 mm. The dose rate range could be extended by increasing the length of the scintillator, but at the cost of spatial or temporal resolution. As the background signal was almost constant in the explored dose rate range, all the scintillators evidenced almost perfectly linear behavior.

The mPSD system studied by Therriault-Proulx et al.\cite{Therriault-mPSD-Brachy-2013} is limited to HDR brachytherapy measurements within 3 cm from the source. The system proposed here is able to accurately perform dose measurements beyond 3 cm with a high collection efficiency.

\section{Conclusions}

In this study, we optimized an mPSD system that can be used clinically in HDR brachytherapy. We found that the scintillation light emission per millimeter of scintillator was more efficient in BCF-10 than in BCF-12 or BCF-60 scintillators. Furthermore, we experimentally determined the appropriate position of each scintillator inside the fiber: the scintillating element with the shorter wavelength should be placed closer to the photodetector, whereas the scintillator with the longer wavelength should be placed distally. In a 2-point mPSD, the most balanced signal was obtained with BCF-10 placed proximally and BCF-60 placed distally. We also evaluated a 3-point mPSD consisting of BCF-10, BCF-12, and BCF-60 scintillators. The best prototype used 3 mm of BCF-10, 6 mm of BCF-12, and 7 mm of BCF-60. Those dimensions were determined not only on the basis of light emission balance, but also with the aim of improving the detector's spatial resolution. Finally, an optimal light collection system was evaluated in HDR brachytherapy simulations. The evaluated mPSD produced minimal deviations in dose rate readings, and analysis of SNR and SBR showed that the detector provided accurate real-time dose measurements.

\begin{acknowledgments}

The present work was supported by the National Sciences and Engineering Research Council of Canada (NSERC) via the NSERC-Elekta Industrial Research Chair grant No. 484144-15 and by a Canadian Foundation for Innovation (CFI) JR Evans Leader Funds \#35633. Haydee Maria Linares Rosales further acknowledges the support from Fonds de Recherche du Qu\'ebec - Nature et Technologies (FRQ-NT) and  by the CREATE Medical Physics Research Training Network grant of the Natural Sciences and Engineering Research Council of Canada (Grant \# 432290).  We also would like to thank Amy Ninetto from the Department of Scientific Publications of the University of Texas MD Anderson Cancer Center for editing our manuscript. 

\end{acknowledgments}

%%%%%%%%%%%%%%%%%%%%%%%%%%%%%%%%%%
% Bibliography section
%\bibliographystyle{ieeetr}
%\bibliography{Bibliography.bib}

\end{document}